
\input phyzzx
\PHYSREV
\unnumberedchapters
\date={April 1993}
\Pubnum={\caps UPR-560-T
 }

\def\to{\rightarrow}
\titlepage
\title{Flat World of Dilatonic  Domain Walls}
\frontpageskip=0.5\medskipamount plus 0.5 fil
\author{  Mirjam Cveti\v c}
\address{Department of Physics\break
University of Pennsylvania\break
Philadelphia, PA 19104--6396\break
}

\abstract{
We study dilatonic  domain walls specific to superstring theory.
 Along with the matter fields and  metric
the  dilaton also  changes its value in the wall background.
We found supersymmetric (extreme) solutions which in general
 interpolate between isolated  superstring vacua with non-equal
 value of the matter
potential; they correspond to the
static, planar  domain walls
with {\it flat} metric in the string (sigma model) frame.
 We point out similarities between the space-time of dilatonic walls
and that of charged dilatonic black holes. We  also comment
 on non-extreme solutions corresponding to expanding bubbles.}

\line{PACS \# 11.17+y, 04.20.-q, 04.65.+e, 11.30.Pb \hfill}
\endpage

\REF\HL{ For an attempt  to study domain walls within the
Jordan-Brans-Dicke theory
see H.-S. La, CTP-TAMU-52/92 hepph/9212041, CTP-TAMU-78/92
hepth/9207202.}

 \REF\VILENKIN {A. Vilenkin, Phys.\ Lett.\ {\bf 133B},  177 (1983).}

\REF\IS{J. Ipser and P. Sikivie, Phys.\ Rev.\  D {\bf 30},
712 (1984).}
\REF\CGSII{M. Cveti\v c, S. Griffies and H. Soleng,
{\it  Non- and Ultra- Domain
Walls and their Global Space-Time}, UPR-546-T/Rev (March 1993).}

\REF\CGRI{ M. Cveti\v c, S. Griffies, and S.-J. Rey, Nucl.\ Phys.\
{\bf B381}, 301 (1992).}

\REF\CG{ M. Cveti\v c and
S. Griffies, Phys.\ Lett.\ B {\bf 285},  27 (1992).}

\REF\CGII{ For a review, see: M. Cveti\v c and
S. Griffies, in  {\em Proc.\ Int.\ Symp.\ on Black Holes,
Membranes and Wormholes\/}, The Woodlands, Texas, January 1992,
edited by S. Kalara and D. Nanopoulos  (World
Scientific, Singapore, in press).}

\REF\CDGS{ M. Cveti{\v{c}}, R. Davis, S. Griffies and H. H. Soleng,
Phys.\ Rev.\ Lett.\ {\bf 70}, 1191 (1993).}

\REF\Gibb{G. W. Gibbons,
DAMTP preprint R-92/38, October 1992,
Nucl.\ Phys. B (in press).}

\REF\CBH{G .Gibbons, Nucl.  Phys. {\bf B207}, 337 (1992);
G. Gibbons and K.
Maeda, Nucl. Phys. {bf B298}, 741 (1988);
D. Garfinkle, G. Horowitz and A.
Strominger, Phys. Rev. {\bf D43}, 3140 (1991), {\bf D45},
3888 (1992){\bf E}.}
\REF\HOR{ For a review   see G. Horowitz, {\it The
Dark Side of String Theory: Black Holes and Black Strings}, USCB-92-32
(October 1992) and references therein.}
\REF\BKT{ V. A. Berezin, V. A. Kuzmin, and I. I. Tkachev,
Phys.\ Lett.\ {\bf 120B}, 91 (1983).}
\REF\DESITTER{ H. Sato, Progr.\ Theor.\ Phys.\ {\bf 76}, 1250 (1986);
S. K. Blau, E. I. Guendelman, and A. H. Guth,
Phys.\ Rev.\ D {\bf 35}, 1747 (1987);
V. A. Berezin, V. A. Kuzmin, and I. I. Tkachev,
Phys.\ Rev.\ D {\bf 36}, 2919 (1987).}
\REF{\CD} {S. Coleman and F. De~Luccia, Phys.\ Rev.\ D {\bf 21},
3305 (1980). }

We address  dilatonic domain wall
solutions\refmark\HL\ which are specific to superstring
theory. In the domain wall
background along with the  matter fields
and  metric
 the dilaton field also changes its value. Such
walls are of particular interest because they correspond to
configurations which  interpolate between isolated
superstring vacua and may thus
shed light on the nature and connectedness  of the superstring vacua.
The primary goal is to present  supersymmetric (extreme) solutions
 which correspond to
static, planar  domain walls in general interpolating between  isolated
 four dimensional  ($4d$) superstring vacua with
non-equal  value of the matter  potential.
 We also comment on  non-extreme walls.

Dilatonic   domain walls  are a generalization
of the ``ordinary''   domain
walls\refmark{\VILENKIN ,\IS}
 in an analogous way as   dilatonic charged black
holes\refmark{ \CBH ,\HOR} are a generalization of
``ordinary'' black holes.
Ordinary domain walls  between vacua of non-equal cosmological constant
fall into three classes:\refmark\CGSII
{\it (i)} extreme  (supersymmetric)  static, planar  domain
 walls,\refmark{\CGRI, \CG, \CGII ,\CDGS,\Gibb}
  {\it (ii)} non-extreme domain
walls (expanding bubbles with an inertial observer  inside the bubble
for  each side of the wall)\refmark{\CGSII ,\IS} and {\it (iii)}
ultra-extreme  walls (expanding
bubbles\refmark{\BKT ,\DESITTER} of false vacuum
decay\refmark\CD).  The energy density $\sigma_{non}^{ultra}$
of the non- [or ultra-]extreme walls
is bound from below [or above]
 by the one $\sigma_{ext}$
of the extreme ones.
Walls  are thus an example\refmark\CGSII\  of
configurations  for which   supersymmetry provides
a   lower   bound for the energy of stable wall
configurations.\Ref\KalGib{ In the black hole context, see
G. W. Gibbons in: {\em Supersymmetry,
Supergravity and Related Topics,\/} edited by F. del Aguila et al.\
(World Scientific, Singapore 1985), p. 147;
R. E. Kallosh, A. D. Linde, T. M. Ort{\' {\i}}n, A. W. Peet,
and A. van Proeyen, Phys.\ Rev.\  D {\bf 46}, 5278 (1992).}
The space-times induced by the walls are non-singular
with non-trivial global structure
and  horizons closely related to the
ones of certain black holes: on the
anti-deSitter [or Minkowski] side of the wall the induced
non-singular space-time is closely related to the ones
 of  the
Reissner-Nordstr\" om  [or Schwarzschild] black holes.
\refmark{\CGSII}\
  The intriguing similarity between
the  space-time of the walls and   the
 one of the corresponding black-holes  reappears  in
the case of dilatonic walls as well.
In the domain wall case the role of the mass ($M$) and
the charge ($Q$) of the black hole is played by the
energy density ($\sigma$)  of the
wall  and the cosmological constant ($\Lambda$)
outside the wall, respectively.

\REF\Witten{E. Witten, Nucl. Phys. {\bf B268}, 373 (1986).}
\REF\WittenII{E. Witten, Phys. Lett. {\bf B155}, 151 (1985).}
\REF\FOOTII{ A superpotential for the dilaton can, however,  be induced
non-perturbatively.
 Also, the  K\" ahler potential
 receives  one-loop corrections due
to  mixed Yang-Mills-$\sigma$ model anomalies. See,
G. Cardoso and   B. Ovrut, Nucl. Phys. {\bf B369}, 351 (1992); J.P.
Derendinger, S Ferrara, C Kounnas and F. Zwirner,
Nucl. Phys. {\bf B372}, 145
(1992).
The above effects should
eventually be included in the  full treatment of
the theory.}

 Potentially phenomenologically viable superstring vacua
 are described by an effective  $4d$ $N=1$ supergravity theory.
The scalar part of the  effective Lagrangian  involves the  metric
$g_{E \mu \nu}$,
the  dilaton $S\equiv e^{-2\phi}+ ia$ (written in this form as
 a scalar part of the chiral superfield),\Ref\FOOTM{Since  we study
the part of the effective Lagrangian which depends   on $S+S^*$ only,
 the Lagrangian
in terms of the linear dilaton
supermultiplet is equivalent to the one in
terms of the chiral dilaton supermultiplet.}
matter fields and gauge fields. In this note we do not include
gauge fields; however, since the dilaton  does couple to  gauge fields
 the study of  charged dilatonic walls is interesting and will be
examined elsewhere. For the sake
of simplicity we take only one (complex) matter field $T$,
a scalar component of a chiral superfield interpolating between
isolated minima of the matter potential.

To all orders in string loops the superpotential $W=W_0(T)$
of the effective Lagrangian of
superstring vacua does not depend on the dilaton,\refmark\Witten\
\ie , it is only  a function of the matter fields.
In the
K\" ahler  potential $K$ the dilaton
couples\refmark{\WittenII ,\FOOTII}\ in a special
way: $K=-\kappa^{-1}\log (S+S^*)+K_0(T,T^*)$.
We  put the imaginary part (axion) of the dilaton  field to
zero ($a=0$) which   turns out  to be  the solution  of field
equations for the dilatonic  domain walls anyway.
The scalar part of the
Lagrangian (in the Einstein frame) is then of the form:
$$
L_E = \sqrt{-g_E}(-{1 \over {2\kappa}}R_E
+\kappa^{-1}g_E^{\mu\nu}\partial_{\mu} \phi
\partial_{\nu}\phi + T_0-{e^{2\phi} \over 2}\tilde V_0)
\eqn\EL
$$
where $T_0=g_E^{\mu\nu}K_{0_{T T^*}}\partial_{\mu} T
\partial_{\nu}T^* $ is the kinetic energy of the matter field, and
$\tilde V_0=e^{\kappa K_0}\times\break
(K_0^{T  T^*}|D_{T}W_0|^{2} -
2\kappa|W_0|^{2})$
 is the part of the potential that  depends on the matter fields,
only.  Here $K_{0_{TT^*}}\equiv \partial_T\partial_{T^*}K_0$
and $D_{T}W_0\equiv
e^{-\kappa K_0} \partial_T (e^{\kappa K_0} W_0)$. We use the space-time
signature $(+---)$  and $\kappa=
8\pi G $. For supersymmetric minima $D_TW_0=0$ and thus $\tilde
V_0=-2\kappa e^{\kappa K_0}|W_0|^2\leq 0$, \ie ,
supersymmetric minima have non-negative  cosmological constant.
The value of the potential $\tilde V_0= -2\kappa e^{\kappa
K_0}|W_0|^2\leq 0$  at a supersymmetric minimum
is different from the corresponding ``ordinary''  $N=1$ supergravity
one $V_0=-3\kappa
 e^{\kappa K_0}|W_0|^2$. The additional
factor  $\kappa e^{\kappa K_0} |W_0|^2$
 is due to  an additional dilaton contribution $e^{\kappa K}
|D_SW|^2 K^{SS^*}\equiv  {{e^{2\phi}}\over2}\times
\kappa e^{\kappa K_0}
|W_0|^2$ to
the total potential in Eq.\EL\ .

A natural frame  to which strings couple is  the string frame,\ie ,
the frame of the sigma model expansion
of the string effective action. In this
case
 $(g_s)_{\mu\nu}=e^{2\phi}g_{\mu\nu}$
the scalar part of the  action is of the
form:
$$
L_s =\sqrt{-g_s}e^{-2\phi}[ -{1 \over {2\kappa}}R_s
-2\kappa^{-1}g_s^{\mu \nu}\partial_{\mu} \phi
\partial_{\nu}\phi+ T_0 -{\tilde V_0\over 2}]
\eqn\LS
$$
\REF\ISR{ W. Israel, Nuovo Cimento {\bf 44B},  1 (1966).}
\REF\FOOTXX{Outside the wall region the same
solutions   can be obtained (M. Cveti\v c and H.
Soleng, unpublished) in the thin wall approximation. Outside the
wall   $|z|>z_0$
the metric Ansatz
(Eq.\metric )  satisfies Einstein's equations: $\partial_zH-H^2/2
=-2(\partial_z\phi)^2, \ \
\partial_zH+H^2=2\alpha_{1,2}^2e^{2\phi}A_E(z)$,
 where $H\equiv \partial_z \ln A_E(z)$, and the dilaton satisfies
 the Euler-Lagrange equation: $H\partial_z\phi+\partial^2_z\phi+
\alpha_{1,2}^2e^{2\phi}A_E=0$. Here $\alpha_{1,2}
=(-\kappa\tilde V_0/2)^{1/2}_{\ \ 1,2}$.
Using matching conditions\refmark\ISR\
across the wall surface one obtains  $\sigma = H_{z_0^+} -H_{z_0^-}$
which is the same as the saturated  Bogomol'nyi bound \energy .
}
\REF\FOOTXXI{One can obtain  related  supersymmetric
 (``no-scale'') domain wall solutions
also  in the so called no-scale supergravity
 models with
$K=-\kappa^{-1}C\log (M+M^*) +K_0(T,T^*)$ ($C=1,2,3$) and    $W=W_0(T)$.
Along with the matter fields $T(z)$  and  metric also
 the field $M(z)$  changes in the wall's background.
Theories with such a form of  scalar interactions
are  realized in  certain
superstring vacua with
$M$'s corresponding to  untwisted moduli.}

 We are searching for planar
(in $(x,y)$ plane), static, supersymmetric dilatonic domain walls
interpolating between isolated supersymmetric
minima  with $\tilde V_{0\ 1,2}\equiv
-2\kappa e^{\kappa K_0}|W_0|^2_{1,2}$
The  metric Ansatz is of the form:
$$ds_E^{2} = A_E(z)(dt^{2} - dz^{2}-dx^{2} - dy^{2}).
\eqn\metric
$$
 and the  scalar field $T(z)$ and the dilaton $\phi(z)$ depend on $z$,
only.
Using a technique of the generalized Nester's form, as developed for
the study of ordinary domain wall configurations
 in Ref.\CGRI , one
 obtains the relation between supersymmetry transformations  and a
Bogomol'nyi bound                                      for the ADM
energy density of the planar domain
wall
configuration:
$$
\sigma - |C|
=
\int_{-\infty}^{\infty}
[-\delta_{\epsilon}\psi^\dagger_i g^{ij} \delta_{\epsilon}\psi_j
+               K_{T T^*}\delta_{\epsilon}\chi^\dagger
\delta_{\epsilon}\chi +
K_{SS^*}\delta_{\epsilon}\eta^\dagger
\delta_{\epsilon}\eta]dz               \geq
0
\eqn\bogbound $$ where $\sigma$ is the energy per unit
area,                                      $C$ is the topological
charge, $g_{ij}$ is the metric of
the space
coordinates.
 $\delta_{\epsilon}\psi_\mu$, $\delta_{\epsilon}\chi$ and
 $\delta_{\epsilon}\eta$
are the supersymmetry variations of the gravitino,
supersymmetric partner of  the matter  field $T$ and  the  dilaton $S$,
 respectively.
For supersymmetric bosonic backgrounds,  one has
$\delta_{\epsilon}\psi_{\mu} = \delta_{\epsilon}\chi =
 \delta_{\epsilon}\eta=0$
 which yields coupled
first order differential equations (Bogomol'nyi equations) for
 the  metric \metric, the   complex matter
 field $T(z)$ and  the dilaton $\phi (z)$
 Ans\" atze:\refmark{\FOOTXX ,\FOOTXXI}
$$\eqalign{Im(\partial_{z}T{D_{T}W_0\over W_0})& = 0,\cr
    \partial_z T& = -\zeta \left({{A_Ee^{2\phi}}\over
2}\right)^{1\over2}e^{\kappa K_0\over
2}|W_0|
K_0^{TT^*}
{D_{{T}^*}\overline{W_0}\over \overline{W_0}},\cr \partial_{z}\log A_E
&=                                                         2
\zeta\kappa\left({{A_Ee^{2\phi}}\over 2}\right)^{1\over2}
 e^{\kappa K_0\over 2}|W_0|
\cr
\partial_{z}\phi& =-\zeta
\kappa \left({{A_Ee^{2\phi}}\over 2}\right)^{1\over2}e^{\kappa K_0\over
2}|W_0| }
\eqn\summary$$

$\zeta$ is either
$+1$ or $-1$
and can  change  sign
when and only when $W$ vanishes.\refmark{\CGRI ,\CG}

When Eqs.\summary\ are satisfied  the (Bogomol'nyi)  bound for
the energy per area $\sigma$ is saturated
by the absolute value of the topological charge $|C|$.
The charge can be unambiguously determined in the
thin wall approximation. Then
in the wall region  ($z\sim z_0$)
the  matter field $T$  varies  rapidly  while the metric  and
 the dilaton  are slowly varying.  We
 normalize $A_E(z_0)= 1$ and chose  the  boundary
condition  $e^{2\phi(z_0)}=1$:\Ref\FOOTIX{
A more general  boundary condition
$e^{2\phi(z_0)}= e^{2\phi_0}$ allows  for a family of  one parameter
solutions with  the energy density
\energy\ determined in terms of  $\alpha_{1,2}\equiv
 \kappa e^{(\phi_0+\kappa K_0/2)}|W_0|_{1,2}$. For $\phi_0> 0$,
$\sigma$ is increased due to an additional dilaton contribution.
 Also, Eq.\flat\  is rewritten
as $ A_s(z)=A_E(z)e^{2\phi(z)}=e^{2\phi_0}$ and
 $A_E(z)= \exp[\int_{z_0}^z\ dz\
\sqrt2\zeta\kappa e^{(\phi_0+\kappa K_0/2)}|W_0|]$.}
$$
\sigma = |C| \equiv \kappa^{-1}\sqrt{ 2}(\alpha_1\pm\alpha_2)
\eqn\energy
$$
Here,
$\alpha_{1,2}\equiv \kappa e^{{\kappa K_0 / 2}} |W_0 |_{1,2}=
(-\kappa\tilde
V_0/2)^{1/2}_{\ \ 1,2}$  where
subscript  1 [or 2] refers to the side of  the wall with
a more [or less]  negative value  for $\tilde V_0$. The signs   $\pm$
correspond to the two classes of the solutions with $W_0$ crossing
zero and
 $W\neq 0$ everywhere, respectively. Note, there are no static walls
with $\tilde V_{0\ 1,2}=0$ on both sides of the wall.

 The energy density of ordinary
supersymmetric  domain walls is of a similar form:\refmark{\CGRI ,\CG}\
$\sigma_{ext} =2\kappa^{-1}(\alpha_1\pm\alpha_2)$
where $\alpha_{1,2}\equiv \kappa e^{{\kappa K_0 \over 2}}| W_0 |_{1,2}
=(-\kappa\tilde
V_0/3)^{1/2}_{\ \ 1,2}$ is defined in
terms of $W_0$ and $K_0$ in the {\it same}
 way as above.
An additional factor  $1/\sqrt 2$  in the case of dilatonic walls is
associated with the dilaton contribution
to  the quantity $ e^{{\kappa K \over 2}}| W_0 |_{1,2}=1/\sqrt 2 \times
e^{{\kappa K_0\over 2}}| W_0 |_{1,2}$. Namely,  the boundary condition
$e^{2\phi(z_0)}=1$ ensures that the effective
cosmological constant on each
side of the wall is by  a factor of $1/2$ less negative,
thus decreasing
the energy density of the wall  by a  factor of $1/\sqrt 2$. There is
a  parallel relation\refmark\HOR\ between  the mass $M$ and the charge
$Q$ for extreme charged (Reissner-N\" ordstrom)
black holes ($M=Q$)  and extreme charged
dilatonic black holes  ($M= Q/\sqrt2$). In the domain wall  case
the role of the charge is
is played by the parameters
$\alpha_{1,2}$  associated with   the value of the matter
 potential at each  minimum.

The first two equations  in \summary\ govern the evolution
of the matter field $T(z)$; the first
one corresponds to the ``geodesic'' equation\refmark\CGRI\
 for the complex $T$
field and is identical to the one of ordinary supersymmetric domain
walls.\refmark\CGRI\ In the limit, $\kappa\to 0$, it reduces to the
constraint that the  geodesic path of $T$  corresponds to $W$ which is
a straight line through the origin.

 The equation for the conformal factor
$A_E(z)$ and the dilaton field (see Eqs.\summary)
 imply that $A_E(z)e^{2\phi(z)}
=const.$ which   with the  boundary conditions $A_E(z_0)=
e^{2\phi (z_0)}=1$ imply:
$$A_s(z)\equiv A_E(z)e^{2\phi(z)}=1
\eqn\flat
$$
Therefore, the metric factor $A_s(z)$ in the string frame
 is {\it flat},
\ie ,  independent of the
value of the matter potential  everywhere  in the
domain wall background. Although  there
is a nontrivial matter potential,
the dilaton field adjusts itself
in the domain wall background in such a way
as to leave the string  metric flat; strings do not ``feel'' the wall.
In addition, the second equation for the
matter field decouples  from the metric
and dilaton equations and
the metric  factor $A_E(z)$ can
be expressed in terms of the matter field as
$A_E(z)= \exp(\int_{z_0}^z\ dz\ \sqrt2\zeta
\kappa e^{\kappa K_0/2}|W_0|)
$.

There are two types of $AdS$ ($\alpha_1\neq 0$) --
$AdS$ ($\alpha_2\neq 0$) walls corresponding to the two  signs
in Eq.\energy . Here $AdS$ refers to the space-time  with the dilaton
modulated  negative cosmological constant.
 The $+$ sign in Eq.\energy\ corresponds to  solution with $W_0$
 traversing zero; in this case
 the  form of the metric outside the wall is $A_E(z)_{1,2}=
e^{-\sqrt2\alpha_{1,2}|z|}$.
The $-$ sign in Eq.\energy\ corresponds to the case with $W_0\neq 0$
everywhere and the  form of the metric  outside the wall is:
$A_E(z)_1=e^{-\sqrt2\alpha_{1}|z|}$ and
$A_E(z)_2
=e^{\sqrt2\alpha_{2}|z|}$.
 As $z\rightarrow \pm \infty$,
 $A_E(z)=e^{-\sqrt 2\alpha_{1}|z|}\to 0$  and thus both  the dilaton
field  and the  curvature blow up in this region.
 However, this  singularity
is an infinite geodesic distance away.
On the other hand, as $|z|\to \infty$,
 $A_E(z)_2=e^{\sqrt2\alpha_2 |z|}\to\infty$, which
   corresponds to  the zero curvature  space-time
  and  is geodesically complete. Note, that  in this region
$e^{2\phi}=e^{-\sqrt 2\alpha_2|z|}\to 0$ (see Eq.\flat ) and thus  the
effective cosmological constant
 $\Lambda =\kappa^{1/2} e^{2\phi}\tilde V_0/2 \to 0$.

In the following we  discuss  a special case:
 $AdS$ ($\alpha_1\neq 0$) -- $M$($
\alpha_2=0$) walls.  $M$ refers to the Minkowski space with zero
cosmological constant.
 In this case the  thin wall
solution (located at $z_0=0$) has the explicit
form :
$$\sigma  =\sqrt 2\kappa^{-1}\alpha_1;\ \  A_E(z)_1=e^{-\sqrt2\alpha_1
z},
 \ z<0;\  \ A_E(z)_2=1,\  z>0. \eqn\adsm$$
where $\alpha_1$ is defined after Eq.\energy .
 For illustrative purposes we also present in
Figure 1 an explicit  finite size wall
  solution   for $T(z)\in {\cal R}$ (solid
line) and  $A_E(z)$ (dashed line).  We
chose an example  with  $W_0=\sqrt\kappa
T^2(T^2-2a^2/\kappa)$, $K_0=TT^*$ and $a^2=0.1$.   The wall
interpolates between $T=0$ and $T\sim a/\sqrt\kappa$ and has a
thickness ${\cal O}(\sqrt\kappa/a^2)$.\Ref\FOOTXXXI{In the case
$a^2\ll 1$, equations \summary\ can be solved explicitly: $T(z)= a\
e^{\tilde z}/(1+e^{2\tilde z})^{1/2}$ and $A_  E(z)=(1+e^{2\tilde
z})^{-a^2/\sqrt 2}\exp\{-a^2/[\sqrt 2(e^{-2\tilde z}+1)]\}$ where
$\tilde z=a^2 z/\sqrt \kappa$.}

The Penrose diagram   for such   walls in the $(z,t)$  plane is given
on Figure 2. The $M$ side ($\alpha_2=0$)
corresponds to  Minkowski space-time
while  the $AdS$
side ($\alpha_1<0$) exhibits singularity
an infinite geodesic distance away.
Note  a formal similarity with the Penrose diagram\refmark{\CBH,\HOR}
for  the
$(r,t)$ plane of the extreme charged dilatonic black hole.

Extreme dilatonic domain walls  are solutions of the
 $4d$ effective superstring action  (evaluated to all
orders in string loops)
 with  isolated  minima of the matter potential preserving
supersymmetry. Eventually,  supersymmetry should be  spontaneously
broken.
 Current proposals
rely on non-perturbatively induced gaugino
condensates (of the hidden gauge
groups)\Ref\DWRS{ L. Ib\~
anez and P. Nilles, Phys. Lett. {\bf 155} 65 (1985); M. Dine, R.
Rohm, N. Seiberg and E. Witten, Phys. Lett. {\bf B156}, 55 (1985).}
which  introduce   new terms  $\propto e^{-c S}$ in the superpotential.
 Here c is  a positive constant
proportional to the beta functions
of the hidden  gauge groups.
The analysis for this case has to be redone; there is a wealth of new
 wall  solutions which need
not be planar and static anymore and will be addressed elsewhere.
Within dilatonic black holes analogous
\REF\GH{R. Gregory and J. Harvey, Enrico Fermi Preprint EFI-92-49  and
J. Horn and G. Horowitz, Santa Barbara Preprint UCSBTH-92-17.}
\REF\Nappi{$2d$ black holes with the
massive dilaton were studied by  M.
McGuigan, C.Nappi, and S. Yost,
Nucl. Phys. {\bf B375}, 421 (1992). }
solutions  with  different dilaton  potentials
has been studied in Refs.~\GH,\Nappi.
 \REF\FOOTVI {Note that other solutions
 $A_E(z)_2=e^{2\beta|z|}, \phi_{\ 2}=0$
and $A_E(z)_2=\cosh{2\beta |z|},
\exp{2\phi}_{\ 2}=\tanh{(\beta |z|)}^{\sqrt 3}$  correspond
to  expanding bubbles
with an inertial observer inevitably
being hit by the bubble, \ie , those are  bubble of
false vacuum decay.}

There is an alternative possibility where  supersymmetry
 is broken spontaneously by the matter
part   of the potential ($\tilde V_0$).
This case is similar to the case of
non-extreme
of charged dilatonic black holes with $M\neq Q/\sqrt 2$.
  Now, the wall need not be static any more.
A convenient way is to write  the metric in the wall's rest frame
and  assuming  that  the ($(2+1)d$)
space-time   internal to the wall is homogeneous,
isotropic and geodesically
complete. The  general form of metric  (compatible with the
constraint that $\sigma>0$) is then  of the form:\refmark\CGSII\ $ds^2=
A_E(z)[dt^2-dz^2-
(\cosh\beta t)^2/\beta^2d\Omega_2^2]$ . It
corresponds to time dependent bubbles
where $\beta$ parametrizes a deviation of
this solution from the supersymmetric one.
As $\beta\rightarrow 0$, the metric reduces to the extreme one (see
Eq.\metric ).

We address  solutions  corresponding to the
$AdS$ -- $M$ expanding bubbles with an inertial observer  {\it inside}
the expanding bubble on each side of the wall.  On
the $M$  side there is a unique solution: $A_E(z)_2=e^{-2\beta |z|}$,
$\phi_{\ 2}=0$
 which is identical to the ordinary   Minkowski  non-extreme wall
 solution.\refmark{\VILENKIN, \IS ,\FOOTVI}
In the rest frame of the wall the metric solution
exhibits a cosmological horizon identical to the one of the
Schwarzschild black hole horizon.
In the inertial  Minkowski coordinates the wall is
expanding for $t>0$.\refmark{\CGSII}
 On the $AdS$ side  the equations can be solved
perturbatively for  $\beta\gg\alpha$:
$A_E(z)_1=\exp{\{-2\beta |z|-\alpha_1^2/(6\beta^2)
[1-\exp{(-2\beta| z|)}]\}}
, \ \
\phi_{\ 1}=\alpha_1^2/(8\beta^2)[1-\exp{(-2\beta| z|)}]
$.
The energy density of the wall
is :
$\sigma =\kappa^{-1}[4\beta+\alpha_1^2/(3\beta)]$. On $AdS$ side
the metric
also exhibits cosmological horizons whose nature is subject to further
investigation. In particular one is interested in the global space-time
structure  on the $AdS$ side of the wall.

We found extreme (supersymmetric) dilatonic domain walls
 specific to isolated $4d$
superstring vacua.  Such walls are static configurations
 with matter fields in general interpolating
between non-equal minima of the matter potential. Everywhere
 in the domain wall background the
 dilaton field adjusts itself in a way as to
leave metric in the string frame flat;
strings do not ``feel'' the wall.
Intriguing similarities between extreme dilatonic
walls and  extreme charged
dilatonic  black holes are pointed out.

I would like to thank H. Soleng for collaboration  at the initial
 stages of the work and  for useful discussions.
 I also benefited from discussions with
R. Davis, S. Griffies, and J. Horne.
The work was supported by U. S. DOE Grant No.\ DOE-EY-76-C-02-3071, and
 NATO Research Grant No. 900-700.
\endpage

{\bf Figure Captions}

{\bf Figure 1.} Solutions for $T(z)\sqrt\kappa/a$ (solid line)
and $A_E(z)$
 as a function of $\tilde
z=a^2z/\sqrt\kappa$ for the example given in the text and
$a^2=0.1$.

{\bf Figure 2.}Penrose diagram in the $(z,t)$ plane for
the finite size  extreme dilatonic domain wall.
The matter potential $\tilde
V_0=0$ for $z>0$  ($M$ region) and
$\tilde V_0<0$ for $z<0$ ($AdS$ region).
 We use the standard
compactified null coordinates
$-\pi\le (u',v')\equiv 2\tan^{-1}(t\pm z)
\le \pi$.  Note the null singularity on the $AdS$ side.

\endpage

\refout

\end